# AUTOMATED INFERENCE SYSTEM FOR END-TO-END DIAGNOSIS OF NETWORK PERFORMANCE ISSUES IN CLIENT-TERMINAL DEVICES

Chathuranga Widanapathirana, Y. Ahmet Şekercioğlu, Milosh V. Ivanovich, Paul G. Fitzpatrick, and Jonathan C. Li

Department of Electrical and Computer Systems Engineering, Monash University, Australia

{chathuranga.widanapathirana, ahmet.sekercioglu, milosh.ivanovich, paul.fitzpatrick, jonathan.li}@monash.edu

## ABSTRACT

*Traditional network diagnosis methods of Client-Terminal Device (CTD) problems tend to be labor-intensive, time consuming, and contribute to increased customer dissatisfaction. In this paper, we propose an automated solution for rapidly diagnose the root causes of network performance issues in CTD. Based on a new intelligent inference technique, we create the Intelligent Automated Client Diagnostic (IACD) system, which only relies on collection of Transmission Control Protocol (TCP) packet traces. Using soft-margin Support Vector Machine (SVM) classifiers, the system (i) distinguishes link problems from client problems and (ii) identifies characteristics unique to the specific fault to report the root cause. The modular design of the system enables support for new access link and fault types. Experimental evaluation demonstrated the capability of the IACD system to distinguish between faulty and healthy links and to diagnose the client faults with 98% accuracy. The system can perform fault diagnosis independent of the user's specific TCP implementation, enabling diagnosis of diverse range of client devices.*

## KEYWORDS

*Automated Diagnosis, Intelligent Inference, TCP, Support Vector Machines, Network Fault.,*

## 1. INTRODUCTION

For the past decade, the most computer network-related developments have focused on improving connection speeds and developing new applications. However, with the demand for improved network speeds, the tolerance for connectivity and performance issues has decreased. With complex communications networks that support many types of Client-Terminal Devices (CTD), traditional methods of performance and fault diagnosis are increasingly inefficient. Diagnosis of network performance problems requires a methodical approach. First, the faulty segment of the network has to be isolated and second, the exact root cause of the problem should be identified. Analysis of packet traces, especially from the Transmission Control Protocol (TCP), is a sophisticated inference based technique used to diagnose complicated network problems in specialized cases [1]. These traces contain artifacts related to behavioral characteristics of network elements that a skilled investigator can use to infer the location and root cause of a network fault. The expertise and resources required for this technique, however, hinders its usability in the conventional fault resolution process of Internet Service Providers (ISPs).

The most common complaint from broadband users is that their "connection speed is too slow" [2]. ISPs typically employ experienced technical staff that continuously monitor and resolve performance issues in servers, backbone, and access links. Consequently, in often cases, the true bottleneck of a user's connection speed is actually the client device [3]. Commonly, this is the result of overly conservative default networking parameters supplied with almost all out-of-the-box operating systems (OS). Correct configuration of these parameters with respect to the access network technology can improve connection speeds and alleviate user dissatisfaction. In practice however, these settings are difficult for novice users to manipulate. Recent studies have found that network data rates reached by novice users are only one-third of those achieved by expert users, a phenomenon commonly referred to as the "Wizard Gap" [4]. Many common performance issues are simple to correct, but difficult to diagnose. As a result, most customer connection issues persist unresolved [2] and many users experience severely degraded network performance even when the networks are underutilized [5, 6]. The Internet2 performance initiative has found that the median bandwidth in their 10 Gb/s backbone in April 2010 was approximately 3.05 Mb/s [7]. Though solutions have been proposed for improving network traffic conditions [8, 9], little attention has been given to solving the bottlenecks or diagnosing faults at the end-user.

Next-generation OSs (e.g. Google Chrome OS) increasingly use cloud computing to deliver all applications to the user device through the web [10]. A recent report on cloud computing by Armbrust et al. [11] identify performance unpredictability and network bottlenecks among the top obstacles for the adoption and growth of cloud computing.

In this paper, we address the aforementioned issues by introducing a new intelligent inference method for diagnosing network problems using TCP packet traces which we call the Intelligent Automated Client Diagnostic (IACD) system. The system (i) relies only on collection of packet traces upon reporting of a problem, and (ii) focuses on identifying CTD faults and misconfigurations. The authors Widanapathirana et al. previously presented a brief overview of the IACD system in IB2COM 2011.

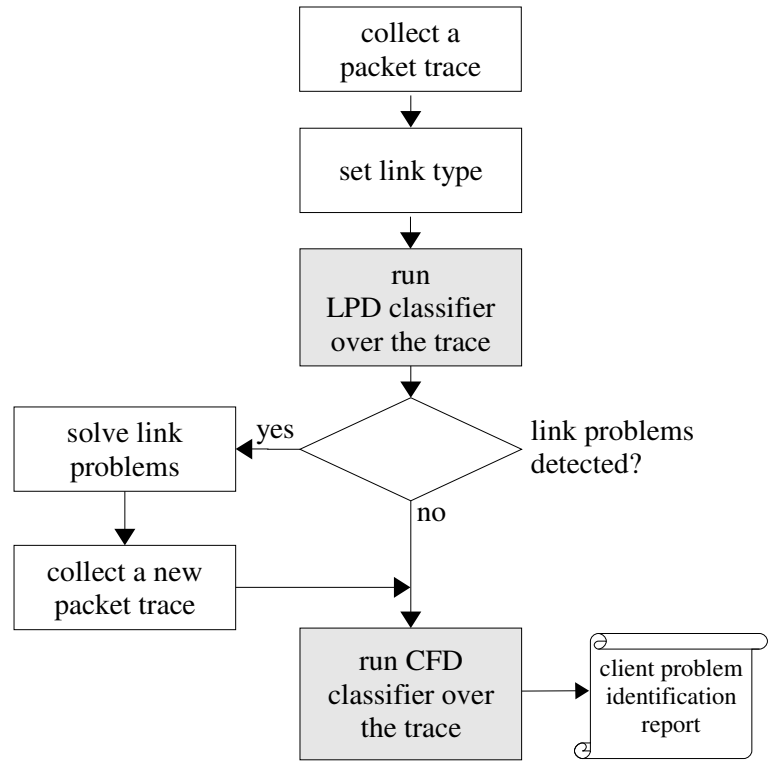

Figure 1: Overview of the operation of the IACD system

## 1.1. Operational Overview of the Intelligent Automated Client Diagnostics (IACD) system

The proposed IACD system is outlined in Figure 1. The system starts by collecting a TCP packet trace of a known stream of data between the client device and the ISP's router. This is initiated by the user through a specially-created web page that activates the trace collection application. This trace is then analysed by the IACD system, which contains two machine learning trained classifiers.

Assuming that the access server is optimized for the connection, performance problems experienced by the end user can be attributed to either the access link or client device. Since the system is designed to diagnose client device problems, a test is first performed to identify whether the performance problem is due to faulty access links.

The TCP packet trace is passed to the first stage of the system, the Link Problem Detection (LPD) classifier that reports whether an access link is operating as expected. If the LPD classifier determines that the access link is faulty, then the link issues should be resolved before attempting to diagnose any client faults. We have left the automatic diagnosis of the access link problems for future work. If the link is identified to be healthy, the packet trace is passed to the Client Fault Diagnostic (CFD) classifier to identify the exact root causes of any problems in the client device.

## 2. BACKGROUND AND RELATED WORK

The use of TCP packet traces for analysis and inference of connection behavior dates back several decades to the introduction of the TCP. The tools and techniques developed since then can be categorized into two main groups: (i) TCP trace visualization tools [12, 13, 14], used to analyze packet streams for organizing and summarizing large amounts of trace data into an easily comprehensible format for expert use, and (ii) TCP-based behavior inference methods using techniques such as packet sequence analysis [15, 16], heuristic analysis [17] and machine learning algorithms [18, 19, 20] to infer connection behavior. These methods have been developed for many applications, including network security [18, 21], system fingerprinting [20, 22], The Internet traffic classification [19, 23], network tomography [24, 15, 16, 25], protocol diagnosis [26, 27, 28], and network diagnosis.

In this paper, we focus our discussion to two categories of diagnostic solutions: (i) rule based behavior inference using TCP traces, and (ii) machine learning based behavior inference using TCP or other event traces.

### 2.1. Rule based behavior inference using TCP traces

The ability to indirectly infer the behavior of protocol layers through the observation of TCP has motivated researchers to create diagnostic systems. The tools "tcpanaly" by Paxton et al. [26] and "TCP Behavior Inference Tool" (TBIT) [29] were early attempts at automated diagnosis of TCP implementations and non-compliance issues, using rule based analysis of TCP packet sequences. Work by Jaiswal et al. [25, 30] on inferring connection characteristics through passive analysis of packet traces used heuristic processes and was later extended to include a more extensive set of rules by Mellia et al. [17, 31, 27].

The project "Web100" [32, 5] focuses on collecting per-connection TCP statistics through kernel instrumentation (KIS) and has received much attention in the research community. It has been used extensively for diagnosing high-speed connectivity issues in projects such as the CERN-Large Hadron Collider and "Visible Human" project [1]. The capability of web100 to capture major protocol events using parameters otherwise hidden from users has been instrumental in developing some more recent tools. NPAD diagnostic servers with "Pathdiag" [33] and "Network Diagnostic Tool" (NDT) [28] use the information extracted using Web100 instrumentation to diagnose connectivity problems of the client systems. These solutions, however, depend on expert-rules for a diagnosis and are unable to detect new faults. With considerable differences among various TCP implementations, rule based systems cannot guarantee an accurate diagnosis unless every exception for a specific rule is considered. Further, the Web100-based tools require dedicated servers or substantial modifications to existing servers. These limitations in the existing rule based solutions have prevented them from being used in commercial networks.

## 2.2. Machine learning based behavior inference

When trained with packet traces representing a specific behavior, supervised machine learning algorithms can identify a similar behavior in test traces. Machine learning based inference of TCP traces have been used in several applications. In recent studies, Dondo et al. [18] used Artificial Neural Networks (ANN), Shon et al. [34] used SVMs, and Kuang et al. [35] used K-Nearest Neighbor (KNN) algorithms to infer network intrusion events using TCP packet traces. With the emergence of a diverse range of internet applications, Internet traffic classification has gained substantial momentum. Hong et al. [36] used Bayesian classifiers for inferring traffic categories from packet traces, and similarly, SVMs have been used by Yuan et al. [19]. Machine learning algorithms have also been used for network tomography applications such as TCP throughput prediction using SVMs [37] and packet loss estimation [38] using Bayesian networks. TCP inference using Bayesian classifiers has been used for remote system fingerprinting by Beverly [20], and in a similar study, Burroni et al. [39] introduced a remote OS identification tool using ANN.

A number of studies have used machine learning for root cause diagnosis of enterprise networks [40, 41, 42, 43, 44], access links [45, 46], home networks [47], and computer systems [48, 42]. These diagnostic tools lack the functionality and generalization required for a broader solution. For example, the decision tree-based "NEVERMIND" [45] is a tool developed only for the diagnosis of ADSL link problems, while the "Netprints" [47] is only used for diagnosing WiFi home network issues. Furthermore, these methods require information such as user requests, event logs, system calls or private network traffic, which demands privileged access.

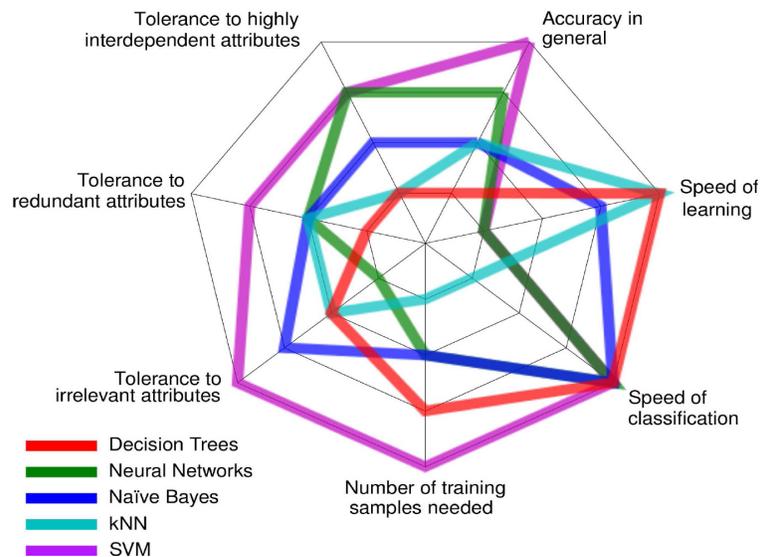

Figure 2: Comparison of machine learning algorithms. Performance of SVM is superior to the others except for speed of learning.

These limitations can be avoided by using an inference based method with an end-to-end TCP connection, independent of the link layer. Our literature survey did not find any comprehensive, scalable, intelligent inference techniques using TCP packet traces for an automatic diagnosis of network performance problems.

The learning algorithm is a critical choice for the intelligent diagnostic system. For this purpose, we first identified candidate algorithms (Decision Trees [49], Artificial Neural Networks [50], Naive Bayes (NB)[51], k-Nearest Neighbor [52], and Support Vector Machines (SVM) [53]), conducted a comparative analysis, and tested their performance with respect to generalization, classification accuracy, number of training samples required, and training speed. Thus, we have chosen the SVM approach because of its superiority in yielding the most accurate results. It also exhibits higher tolerance and requires fewer training samples. The comparative performance of the methods is shown in Figure 2.

## 3. IACD SYSTEM CLASSIFIERS

### 3.1. Link Problem Detection (LPD) classifier

The LPD classifier detects the artifact patterns which exist only when a link's performance degrades from the expected baseline. We define an access link performing at the expected baseline as a healthy access link and a link with degraded performance as a faulty access link. The performance expectation of a healthy link is network-specific. The operator has the freedom to train several LPD classifier modules, each trained for a specific link type and baseline performance (e.g. 24Mb/s or 12Mb/s DSL link, 14Mb/s HSDPA link, 54 Mb/s 802.11g link with 1% packet loss or 5% packet loss).

The problem presented to the LPD classifier conforms to a binary classification problem with two outcomes, either a faulty or a healthy link. The design, as shown in Figure 3 has two phases: first, the training phase creates an appropriate classifier model using two sets of trace samples from faulty and healthy links. The training phase includes signature extraction, data pre-processing and feature selection before pattern classifier training. Second, the diagnostic phase uses the trained classifier model to determine the artifacts hidden in an undiagnosed trace.

After pre-processing and feature selection, the training data set $\Theta_{lpd}$ of $n$ instances is in the form

$$\Theta_{lpd} = \{(\mathbf{x}_i, y_i) \mid \mathbf{x}_i \eth \Re^m, y_i \eth \{+1, -1\}\}_{i=1}^{n} \tag{1}$$

with $x_i$ being an m-dimensional feature vector and class label $y_i$, either +1 for the faulty link or −1 for the healthy link, to which each $x_i$ belongs. For example, a sample trace ($i$=1) from a faulty link, with four features ($m$=4) is denoted by $\{0.5, 0.03, 0, 0.99, +1\}_{i=1}$ (e.g. 1).

We have chosen the L2 soft-margin SVMs [54, 55] with kernel mapping to model the best non-linear separating hypersurface between the faulty class and the healthy class. For an $m$-dimensional input feature vector, the resultant class boundary is an $m$-dimensional hypersurface that separates the two classes with the maximum margin.

For the data set given in (eq. 1), a linear decision function

$$D(X) = \mathbf{w}^T X_i + b \text{ for } i = 1, ..., n, \tag{2}$$

where, $w$ is the $m$-dimensional weight vector. The optimum hyperplane is found by minimizing

$$Q(\mathbf{w}, b, \xi) = \frac{1}{2}\mathbf{w}^T\mathbf{w} + \frac{C}{2}\sum_{i=1}^{n}\xi_i^p \tag{3}$$

with respect to *w*, *b* and $\xi$, subject to the inequality constraints:

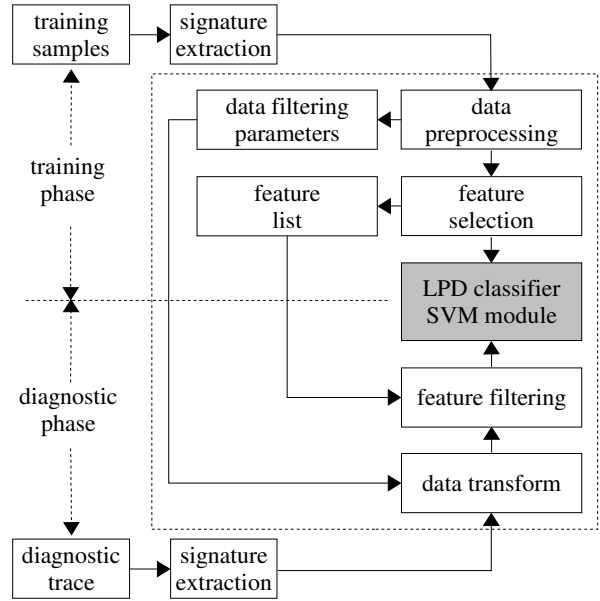

Figure 3: LPD classifier design

$$y_i(\mathbf{w}^T\mathbf{x}_i + b) \geq 1 - \xi_i \text{ for } i = 1,...,n \text{ and } \xi_i \geq 0 \qquad (4)$$

where, nonnegative $\xi$ is called the slack variable and allows a degree of inseparability between the two classes and $p = 2$ for L2 soft-margin SVM.

This conforms into a convex quadratic programming problem (QP) [56] and solved after converting the constrained problem given by (3) and (4) into an unconstrained problem

$$L(\mathbf{w},b,\alpha,\xi) = \frac{1}{2}\mathbf{w}^T\mathbf{w} + \frac{C}{2}\sum_{i=1}^{n}\xi^2 - \sum_{i=1}^{n}\alpha_i\left(y_i(\mathbf{w}^T x_i + b) - 1\right) + \xi_i) \qquad (5)$$

where $\alpha_i$ ($\geq 0$) are the Lagrange multipliers introduced to enforce the positivity of the $\xi$. The optimal saddle point ($w_0$, $b_0$, $\alpha_0$, $\xi_0$) is found where $L$ is minimized with respect to *w, b, and* $\xi_i$ and maximized with respect to $\alpha_i$ ($\geq 0$) following Karush-Kuhn-Tucker (KKT) [57] conditions. This is called the training of SVM.

However, to enhance the separability of the linearly inseparable data, using the non-linear vector function $g(\mathbf{x}) = g_1(\mathbf{x}),...,g_l(\mathbf{x})$, the original *m*-dimensional input vector *x* is mapped into the *l*-dimensional feature space. The linear decision function for the obtained *l*-dimensional feature space is given by (2).

$$D(\mathbf{x}) = \mathbf{w}^T g(\mathbf{x}) + b \qquad (6)$$

where *w* is the *l*-dimensional weight vector, *b* is the bias term. According to the Hilbert Schmidt theory, the mapping function g(x) that maps x into the dot-product feature space satisfies

$$K(\mathbf{x},\mathbf{x}_i) = g(\mathbf{x})^T g(\mathbf{x}_i) \qquad (7)$$

where $K(x,x_i)$ is called the kernel function. The kernel function avoids the actual mapping $g(x)$ and directly calculates the scalar products $g(x)^T g(x_i)$ in the input space. We cross validated and analysed the performance of the classifier for multiple kernel functions (eq. 8) to select the best suitable kernel for the classification problem.

$$K(\mathbf{x}, \mathbf{x}_i) = (\mathbf{x}^T \cdot \mathbf{x}_i) \qquad \text{Linear kernel,} \qquad (8a)$$

$$K(\mathbf{x}, \mathbf{x}_i) = (\mathbf{x}^T \cdot \mathbf{x}_i + 1)^2 \qquad \text{Quadratic kernel,} \qquad (8b)$$

$$K(\mathbf{x}, \mathbf{x}_i) = (\mathbf{x}^T \cdot \mathbf{x}_i + 1)^3 \qquad 3^{\text{rd}} \text{ degree polynomial kernel,} \qquad (8c)$$

$$K(\mathbf{x}, \mathbf{x}_i) = \exp(-\parallel \mathbf{x} - \mathbf{x}_i \parallel^2) / 2\sigma^2 \qquad \text{Gaussian RBF kernel,} \qquad (8d)$$

The LPD classifier is capable of detecting faulty links, even if both link and client faults simultaneously cause connection problems. However, this task is challenging because artifacts created by the client faults either (i) mask those artifacts from faulty links, or (ii) create false positives as link problems. To create a robust LPD classifier model, the training data should contain traces collected with faulty as well as healthy clients. Also, a robust feature selection method to identify the unique features that enable the detection of a faulty path regardless of a client behavior should be used (Section 4.5).

The use of a mechanism to detect faulty links before diagnosing a client ensures that, traces sent through the CFD classifier do not contain any faulty link artifacts. This simplifies the design of the CFD classifier and improves classification accuracy.

## 3.2. Client Fault Diagnostic (CFD) classifier

The first stage of the IACD system ensures that the access link is not causing the connection problem. The second stage, Client Fault Diagnostic (CFD) classifier identifies the specific types of client faults, if any, that cause the performance problem.

Choosing between binary classification and multi-class classification is an important design choice for the CFD classifier. There are many studies [58, 59] that compare these two types of classification methods for different applications, each with their own merits.

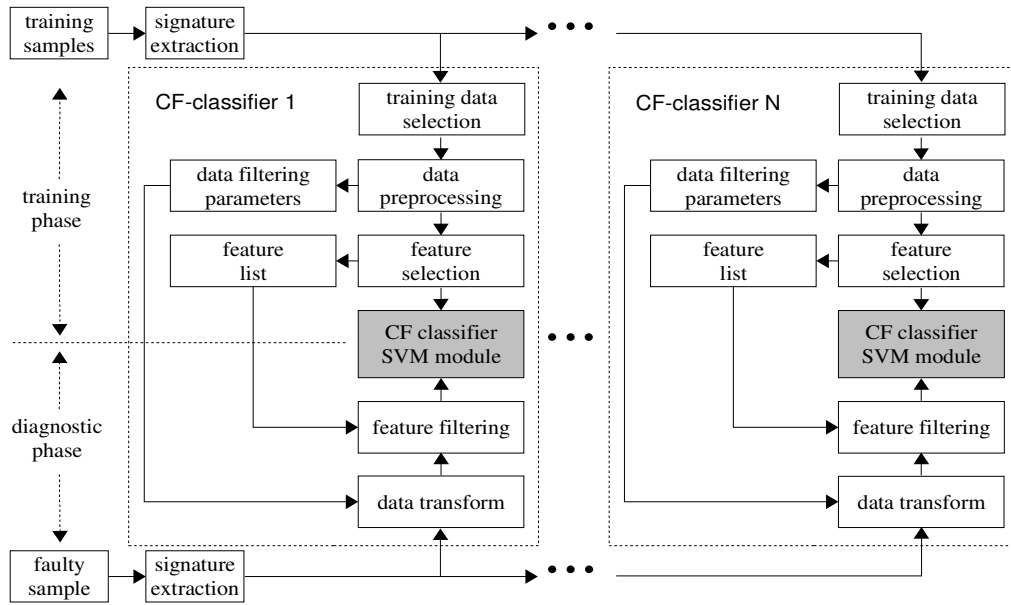

Figure 4: CFD classifier design for the IACD system.

In our CFD classifier design, we opted to use a parallel network of binary classifier modules (CF-classifiers), each trained to diagnose a single fault. This arrangement (Figure 4) collectively performs a multi-class classification. A network of binary classifiers were chosen over a single multi-class classifier because of the (i) flexibility to continually add new diagnostic capabilities to the IACD system, without having to change the basic algorithm or retrain the complete system, (ii) freedom to choose classifier parameters optimized to detect a specific type of artifact independently for each module instead of a common set of classifier parameters for all the faults and, (iii) parallelism which can reduce the time of classification, especially when the classes increase in number.

A client causing the connection problems is defined as a faulty client and a client that does not is defined as a healthy client. The training samples are stored in a trace database $\Theta_{cfd}$ after the signature extraction process.

$$\Theta_{cfd} = \{(\mathbf{x}_i, y_i) | \mathbf{x}_i \grave{o} \Re^m, y_i \grave{o} \{cf_0, cf_1, cf_2, ...., cf_p\}\}_{i=1}^n \qquad (9)$$

where $x_i$ is the $m$-dimensional feature vector and $y_i$ is the class label. The class label $y_i = cf_0$ for a healthy client and $y_i = cf_{1,} cf_{2, ...,} cf_p$ for $p$ types of different client faults. Each module then selects the training data subset $\Theta_{cf,j}$ with traces labeled as $cf_j$ for the faulty class and $cf_0$ labeled traces in the healthy class for training the $j^{th}$ binary CF-classifier as in eq 10c.

$$\Theta_{cf,j} = \{(\mathbf{x}_i, y_i) | \mathbf{x}_i \grave{o} \Re^m, y_i \grave{o} \{cf_0, cf_j\}\}_{i=1}^n, \qquad (10a)$$

$$y_i \equiv cf_0 \equiv -1, \qquad (10b)$$

$$y_p \equiv cf_j \equiv +1 \text{ for } j = 1, ..., p. \qquad (10c)$$

Then, each module independently process data, improving the class coherency and selects the unique feature subset (artifacts) that separates the two classes. This feature subset is then sent to the pattern classifier module to model the classifier boundaries. Each CF-classifier module in CFD classifier uses a L2 soft-margin SVMs for pattern classification (as given in equations (2)) similar to LPD classifier design).

# 4. CLASSIFIER TRAINING

## 4.1. Data collection

Training samples are collected either from a controlled test bed or from a service provider's network. Laboratory test beds are preferred, as different connection problems can be accurately emulated in a well-regulated environment. However, the proposed trace collection technique can be easily implemented in operator networks when the network elements and conditions are not reproducible in the laboratory. The diagnostic accuracy of the classifier is highly dependent on the consistency and accuracy of the artifacts collected. Using standard packet capture libraries, our application captures two packet traces; (i) a client download from the server, at the client, and (ii) a client upload to the server, at the server, upon an online request by the user. Both traces are captured as bi-directional packet flows to ensure most connection details are captured. An incompressible file of size 100MB is used for a longer connection time and a lengthier packet stream. Constant file size and large file transfers generate more consistent and accurate fault signatures compared to a short-lived connection.

## 4.2. Trace signature creation

Two collected packet traces are analysed individually, and then the extracted trace characteristics are combined to form an $m$-dimensional feature vector $\mathbf{x}_i$ which contains an accurate representation of the connection. The feature vector $\mathbf{x}_i$, combined with the class label $y_i$ is called the "signature"of the $i^{th}$ instance. We have developed a signature extraction technique based on "tcptrace" [12], an open source trace visualization tool. Our technique extracts 140 different statistical parameters for each trace, which forms a combined total of 280 parameters for each signature. For example, a raw feature vector from a faulty link before any data pre-processing looks like {1249256, 295, 0, 32, 39, 1, 1,..., FAULTY} (e.g. 2). The statistical trace characterization technique transforms a packet stream into a data vector preserving the fault artifacts. The connection characteristics of the TCP trace are accurately encapsulated in the data vector as we collect an extensive set of statistical parameters.

To create the *preliminary signature database ($\Theta_{psd}$)*, the collected traces are then combined and grouped following the class labels.

The signatures are unique, even within the same class as later shown in Figure 5 and 6. However, for each type of fault class, there exists a subset of features with common values, which are specific for that class. This unique subset of features forms the artifact.

### 4.3 Data pre-processing

The raw feature vectors in $\Theta_{psd}$ need further processing before being used for classifier training. This step, called data pre-processing, improves the overall classification accuracy by enhancing data coherency and consistency within the classes.

First, categorical attributes such as the class labels FAULTY and HEALTHY are converted to numeric data, i.e. +1 for the faulty class and -1 for the healthy class. The contribution of each feature for the classification process depends on its variability relative to other features. If one input has a range of 0 to 1, while another input has a range of 0 to 1,000,000, then the greater numeric range can dominate the smaller [60]. To avoid such inaccuracies, the training data set is shifted and re-scaled along each feature as the second part of pre-processing. Data re-scaling also avoids numerical difficulties during the calculation, especially when working with large values. Each feature is linearly scaled to fit in the range 0-1. This process transforms a raw trace signature (e.g 2) into the form {0.20, 0.82, 0, 0.35, 0.90, 1, 1, ..., +1} (e.g. 3). The resultant database is called the *scaled signature database ($\Theta_{ssd}$)*.

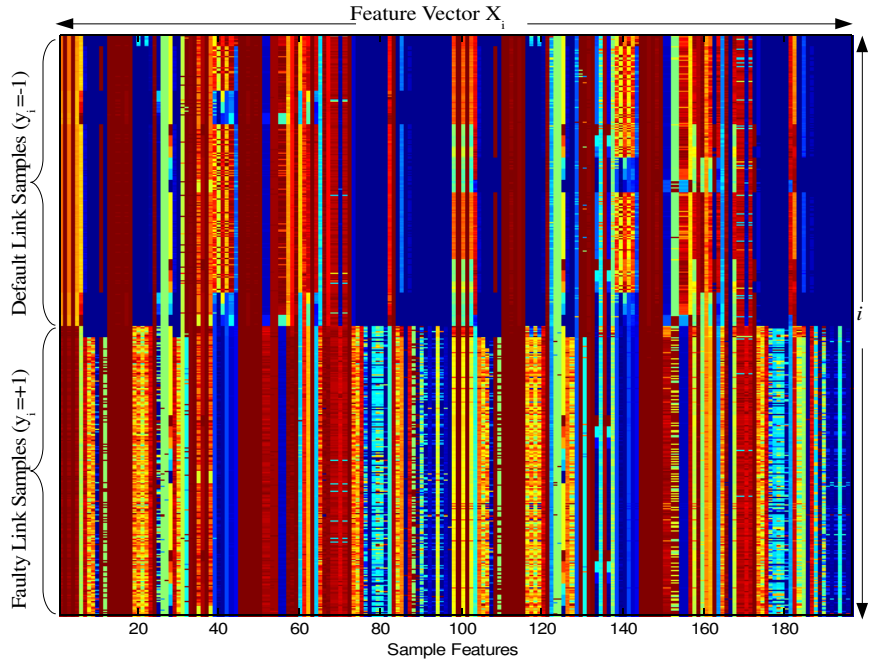

Figure 5: LPD classifier signature database $\Theta_{lpd\_ssd}$, for comparison of faulty and healthy link trace characteristics.

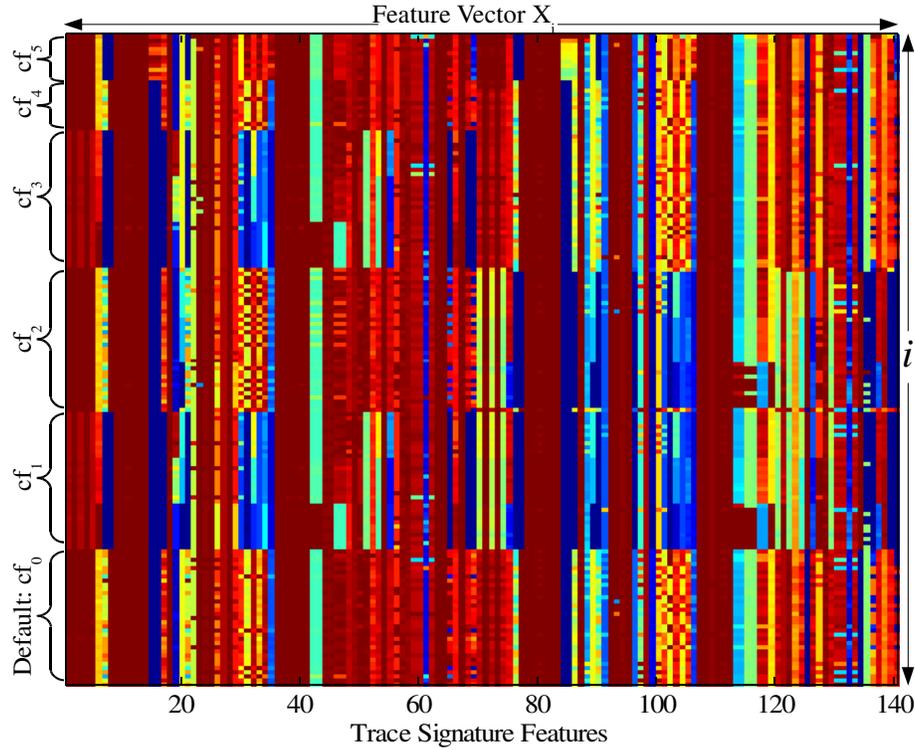

Figure 6: CFD classifier signature database $\Theta_{cfd\_ssd}$, for comparison of client fault classes $cf_i$

## 4.4 Hidden trace artifacts

Figures 5 and 6 show the standardized trace databases $\Theta_{lpd\_ssd}$ and $\Theta_{cfd\_ssd}$ used in training the IACD system. The $i^{th}$ row represent the feature vector $\boldsymbol{x}_i$ of the $i^{th}$ trace sample, colour mapped to RGB space for easy visualization of signature characteristics. Null features have been removed for clarity. Figure 5 shows samples from two classes, faulty path ($y_i = -1$) and healthy path ($y_i = +1$) (see equation (1)). Figure 6 contains samples from multiple client fault classes ($y_i = cf_1, ..., cf_5$) and the healthy client ($y_i = cf_0$) (see equation (9)). Figures 5 and 6 show that the signature extraction process creates unique signatures for every TCP packet trace, even within the same class, preserving the connection characteristics.

In Figure 5, some feature values (columns) behave sporadically (such as features 40-45, 60-65, 160-165 in Figure 5), and provide no usable information to the classifier. However, we can identify a feature subset (features 1-5, 19-22, 115-120, 175-180 in Figure 5) that clearly separates the faulty class from the healthy class. Using these artifacts as a visual guide, the faulty and healthy access links in Figure 5 can be distinguished.

Similarly, Figure 6 shows multiple client fault classes, $cf_i$. The signatures of different fault classes exhibit clear differences compared with the healthy client and are more subtle compared to $\Theta_{lpd}$. When trained, the IACD system automatically identifies and classifies a trace and produces the visually comprehensible classes shown in Figures 5 and 6.

## 4.5 Feature selection

Although the signature format is identical in every sample, only a particular subset of features contributes to the artifact. Unnecessary features increase the computational complexity [61], create overfitting of classifier boundaries [62], and reduce classification accuracy [63]. Therefore, insignificant features should be removed from the training data. In this work, we use an automated feature selection method to select the best suitable feature subset for a particular classifier.

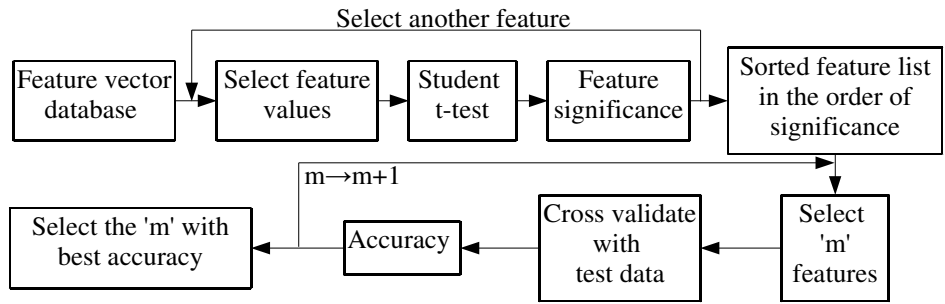

Figure 7: Hybrid feature selection technique for isolating the best feature subset of the artifact.

There are two main categories of feature selection algorithms: (i) filters, which use statistical characteristics of features, and (ii) wrappers, which cross-validate all variations of feature subsets to select the best set. Wrappers are considered to perform better than filters, but are computationally expensive. Our proposed method, as shown in Figure 7, similar to work discussed by Xing et al. [64] and Das et al. [65], follows a hybrid approach (between filter and wrapper) for isolating the best feature subsets. We first use a filter technique, Student's t-test (two-sample t-test) (implemented similar to [66]), to assess the significance of every feature for separating the two classes. Next, the features sorted in the order of significance are cross-validated by incrementing the number of features selected for each class (wrap per technique) against test data to identify the best number of features required for each classifier. Student's t-test is a common statistical data analysis procedure for hypothesis testing, and determines whether two independent populations have different mean values for a specified extent. The feature selection process reduces the m-dimensional feature vector in (1) to $q$-dimensions, where the combination of q features creates the artifact. This process creates a new database, the *optimum signature database* $(\Theta_{osd})$.

## 5. IACD SYSTEM PERFORMANCE

### 5.1 Network emulation

Since we do not have access to operator networks, we used data collected in a network test bed, shown in Figure 8, which emulated an access link, client computer and the access server. The client and server ran on Linux 2.6.32 systems (with Ubuntu distribution), capable of running multiple TCP variants. The access link was emulated using a network emulator, dummynet [67] on FreeBSD 7.3. Each box was connected using full-duplex, 1000 Mb/s cat5e ethernet. Different client and link conditions were emulated using the Linux and dummynet parameter configurations. Then the traces were captured using the technique discussed previously.

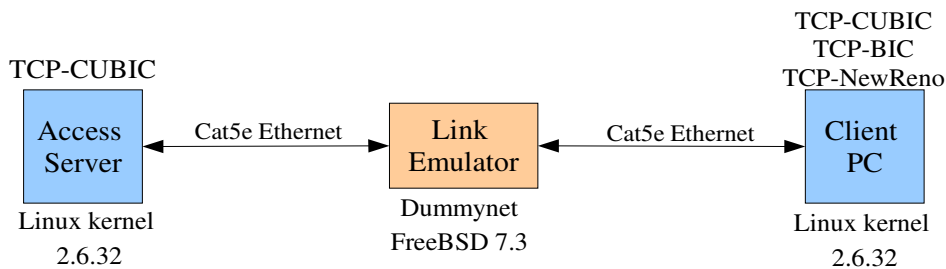

Figure 8: Network testbed for training and testing data collection

## 5.2 Experiment Criteria

Experiments included analyzing the performance of the IACD system with one LPD classifier (for a single type of access link), and CFD classifier with a network of four CF-classifier modules. The experimental setup emulated a full duplex wired access link with a 80 Mb/s bandwidth, 10ms delay with no packet losses and no packet reordering as the healthy link. Faulty links were emulated by inducing packet losses (from 1% up to 10%) and increased delays (from 15ms up to 100ms). Both the server and the healthy client (Linux 2.6.32) had a protocol stack optimized for the healthy link.

For client faults, we emulated the disabled Selective Acknowledgement (SACK) option and the disabled Duplicate Selective Acknowledgement (D-SACK) option, which have been found to cause performance issues in the high-bandwidth connections [68, 69, 70, 71]. Socket buffer limitations, another common and hard to diagnose performance bottleneck [1, 72], were emulated by creating insufficient read buffers and write buffers at the client as two separate cases. Multiple, simultaneous client faults were emulated by creating both read and write socket buffer limitations at the same time. All buffer limitations were emulated using three buffer levels to collect traces from a range of possible scenarios.

Figure 9 shows the data sets used for LPD and CFD classifier training and testing. For training data, both the server and client were limited to run TCP-CUBIC [73], with only 11 traces per each fault class being collected to re-create the worst case practical limitations. To analyze the system performance, we collected four testing data sets as follows: (i) the same data set used in training and, separately collected sets similar to Figure 9 with (ii) TCP-CUBIC client, (iii) TCP-BIC [74] client, (iv) TCP-NewReno [75] client. From other TCP variants, the data sets (iii) and (iv) were collected to evaluate the TCP agnostic properties of the system. An additional data set over a healthy link was collected with clients suffering from multiple simultaneous buffer limitations.

## 5.3 Diagnostic performance of LPD classifier

The training database for LPD classifier $\Theta_{lpd}$, contained two classes, faulty and healthy links (see Figure 5), both of which included different client behaviors. We used 100 traces per class (200 total data sets), each with 280 features, for training the LPD classifier.

The proposed feature selection technique in Section 4.5 requires cross-validation before selecting the best feature subset. Although we cross-validated a number of feature subsets, our analysis is limited to two subsets of 75 and 25 sorted features. Figure 10 shows the two $\Theta_{lpd\_osd}$ databases, where Figure (a) has 75 features, and Figure (b) has 25.

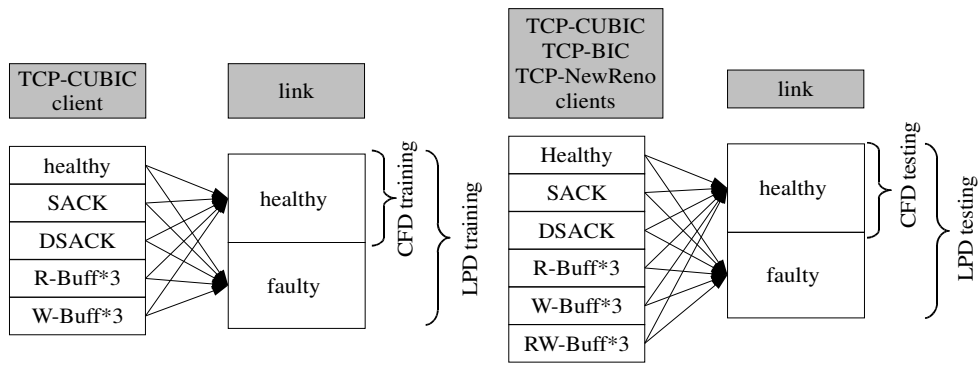

Figure 9: Training and testing data sets for LPD and CFD classifiers. Combinations of client and link conditions are emulated to collect the traces of a variety of possible scenarios.

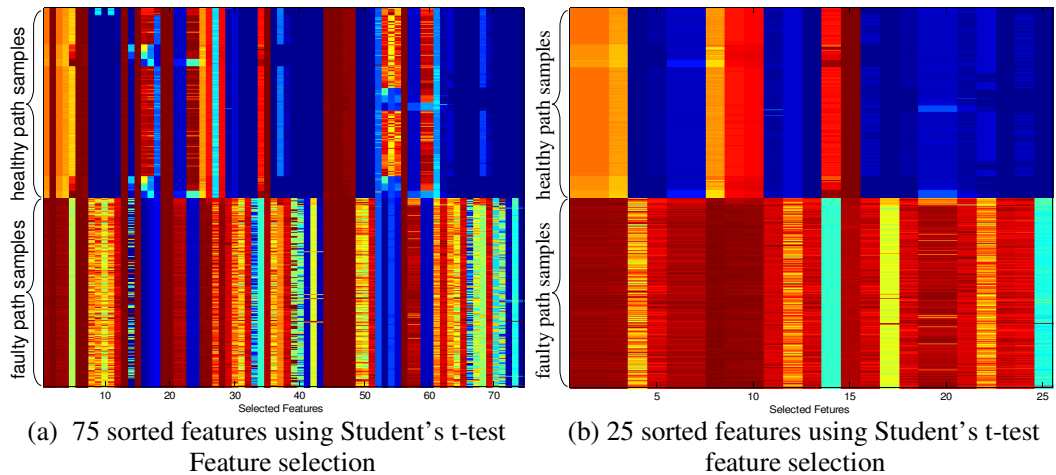

(a) 75 sorted features using Student's t-test
Feature selection

(b) 25 sorted features using Student's t-test
feature selection

Figure 10: Comparison of trace databases after feature selection.

When compared to Figure 5, both feature limited databases show a clearer separation between the two classes. The feature selection technique has reduced the dimensionality of the problem by 73% and 91% with 75 and 25 feature subsets, respectively.

From the kernels in (8), the quadratic kernel was chosen for this particular classifier by cross-validation. Quadratic programming (QP) optimization with 1000 maximum iterations was used to solve (5).

Each of the four testing data sets consisted of 264 traces (132 faulty, 132 healthy link), collected with healthy and faulty clients. The 75 features used for training the classifier create a more complicated class boundary compared to 25 features. This phenomenon is called boundary over-fitting. With over-fitted boundaries, even a small deviation of the data (vector) at the boundary can cause a misclassification. For the first and second test cases (TCP-CUBIC traces), the behavior and artifacts of traces were extremely similar, if not identical to the training traces. The Table 1 shows the diagnostic performance of the LPD classifier for both the 75 and 25 feature subsets. For the first test using the training instances, the classification accuracies were 100%. The accuracy remained at 100% during the second test when the previously unseen TCP-CUBIC data set was used. In these two cases, the lack of outlier samples resulted in high classification accuracy, even with 75 features and an over-fitted boundary. However, the over-fitted boundaries resulted in classification errors of 0.175% for TCP-BIC, and 2.622% for TCP-NewReno data when the artifacts subtly deviated from those of TCP-CUBIC. When the dimensionality was reduced to 25 features, the LPD classifier created a more generalized boundary capable of compensating for artifact variations. As a result, the classifier was highly successful in separating the two classes with 100% accuracy for both the TCP-BIC and TCP-NewReno cases.

Table 1. Classification accuracy of LPD classifier for detecting faulty links. The classifier was successful in identifying the faulty links, even in the cases of faulty clients and TCP variants.

| Link Problem Detection Accuracies | | |
|---|---|---|
| **Trace Samples** | **75features** | **25 features** |
| TCP-CUBIC training set | 100% | 100% |
| TCP-CUBIC testing set | 100% | 100% |
| TCP-BIC testing set | 99.825% | 100% |
| TCP-NewReno testing set | 97.378% | 100% |

## 5.4 Diagnostic performance of CFD classifier

For testing the performance of our CFD classifier design, we created a parallel network of four CF-classifiers, each tasked with diagnosing a commonly found client issue: disabled SACK option (CF-Classifier 1), disabled D-SACK option (CF-Classifier 2), insufficient read buffer (CF-Classifier 3) and insufficient write buffer (CF-Classifier 4). Each training data set had two classes; a healthy client class, and the specific faulty client class, all collected over a healthy access link.

The choice of classifier module parameters has a significant impact on the performance. Table 2 shows the parameters chosen for each of the four CF-classifiers. The cross-validation technique for selecting the parameters and feature subset considered not only the individual classifier accuracy, but also the possible false positives. QP optimization used a maximum of 2000 iterations instead of the 1000 used for LPD classifier training, since the optimization problems were more complex in CF-classifiers with fewer detectable artifacts.

The Table 3 shows the diagnostic accuracy of the CFD classifier, which considers the collective output of the CF-classifier network. When tested with the CUBIC training and testing data sets, the system was capable of diagnosing the client's disabled SACK option, disabled D-SACK option, read buffer limitation and write buffer limitations with high accuracy. Similarly, when tested with TCP-BIC and TCP-NewReno, variants not used during the training phase, the four client faults were diagnosed with 100% accuracy. These results demonstrated the TCP-independent nature of the proposed CFD classifier design.

The healthy clients were identified with 94.81% and 93.5% accuracy during the first two tests of TCP-CUBIC train and test data sets. When samples from healthy clients with TCP-BIC and TCP-NewReno were tested, the detection accuracies were at 92.10% and 91.71%, marginally lower than the other cases. This is due to the slightly higher tendency of obtaining a false positive in at least one of the CF-classifiers by healthy clients' traces compared to other samples. When presented with traces taken from clients with simultaneous read and write buffer deficiencies, CF-classifier 3 and CF-classifier 4 were capable of independently identifying the faults from the trace. This capability led to a collective diagnostic accuracy of 96.97%, 96.90% and 100% for CUBIC, BIC and NewReno data sets, respectively.

## 5.5 IACD system characteristics

For the root cause diagnosis of client performance problems, the proposed IACD system offers many advantages over the other available trace inference methods.

• The system offers a fully-automated, comprehensive framework which is extendible to diagnose a diverse range of faults, contrary to the limited capabilities of other tools.

Table 2. SVM parameters used in each CF-classifier module of the CFD classifier.

| Non-linear SVM Parameters | | | | |
|---|---|---|---|---|
| | CF-Classifier 1 SACK problem | CF-Classifier 2 DSACK problem | CF-Classifier 3 Insufficient Read Buffer | CF-Classifier 4 Insufficient Write Buffer |
| **Kernel** | Linear | RBF | cubic polynomial | RBF |
| **Features** | 12 | 32 | 24 | 16 |

Table 3. Diagnostic accuracy of the CFD classifier, derived from the collective output of the CF-classifier network.

| Trace Samples | Client Fault Diagnostic Accuracies | | |
|---|---|---|---|
| | SACK problem | DSACK problem | Insufficient Write Buffer |
| CUBIC training set | 100% | 100% | 100% |
| CUBIC testing set | 100% | 100% | 93.94% |
| BIC testing set | 100% | 100% | 100% |
| New-Reno testing set | 100% | 100% | 100% |
| | Insufficient Read Buffer | Insufficient Read-Write Buffer | Default Client |
| CUBIC training set | 100% | | 94.81% |
| CUBIC testing set | 96.36% | 96.97% | 93.50% |
| BIC testing set | 100% | 96.90% | 92.10% |
| New-Reno testing set | 100% | 100% | 91.71% |

- Diagnostic capability of the system evolves with the diversity of the fault signature databases, instead of the inference algorithm. Users can collaborate to create common signature repositories, encompassing a wide range of faults, networks, and client platforms. Most rule based systems lack the generality to operate effectively in a dynamic environment.

- The system relies solely on packet traces collected at end-points and can be implemented as an application. This provides flexibility for the operator to deploy the IACD system at any desired network location.

- End-user systems can be diagnosed without remotely accessing or physically logging on to the systems; a capability unavailable in many network diagnostic tools.

- The proposed technique avoids both the idiosyncrasies of individual TCP implementation and the usage of TCP flags as an information source. Instead, the connections are characterized using per-connection statistics of a signature independent of the TCP variant and the negotiated flags.

- Although the system is designed to diagnose client computers from the edge of the operator's network, the same system can be used for diagnosing intermediate nodes in the network by deploying a trace collection module in a neighboring node and training with suitable data.

## 5. CONCLUSION

In this work, we have proposed and evaluated the IACD system, an automated CTD diagnostic system that uses an intelligent inference based approach of TCP packet traces to identify artifacts created by various faults. The system consists of two cascading levels of classifiers: (i) the LPD classifier, tasked with first filtering out whether the connection performance problem is caused by link faults or otherwise, and (ii) the CFD classifier, tasked with diagnosing the specific client faults that cause the connection problem. The LPD classifier uses a single module that incorporates signature extraction, data-preprocessing, feature selection and a soft-margin binary SVM for pattern classification. The CFD classifier performs a complex multi class classification of client faults using a parallel network of CF-classifiers. The modular design of the CFD classifier offers extendibility to diagnose new faults by training CF-classifier modules independently.

We evaluated the system by diagnosing four types of common client problems with various TCP implementations. Furthermore, we analyzed system performance when in the absence of any client faults as compared to multiple simultaneous faults. Our results show that the LPD classifier can effectively identify and separate out the link problems, without being affected by the client behavior and TCP type. The CFD classifier results show that, with a small number of training samples, CF-classifier modules collectively produce high diagnostic accuracy in all tested scenarios, including clients with different faults, TCP variants, default clients and multiple faults.

Our proposed IACD system provides a framework for an accurate diagnostic system that is effective for a large array of client platforms, easy to deploy, and extendible in diagnostic capability. To our knowledge, the IACD system is the first to use automated inference of TCP packet traces using SVMs for diagnosing the root causes of the network performance issues.

**Authors**

**Chathuranga H. Widanapathirana** is a Ph.D candidate in the Department of Electrical and Computer Systems  Engineering at Monash University, Melbourne, Australia. He received his B.Eng degree in Electronics with a  major in Telecommunication Engineering at Multimedia University (MMU), Malaysia in 2009. His research interests include distributed cooperative networks, cloud computing networks, automated machine learning systems, and end-user self-diagnostic services in multiuser networks.

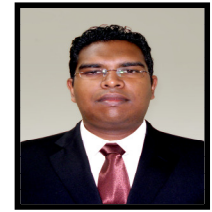

**Y. Ahmet Şekercioglu** is a member of the academic staff at the Department of Electrical and Computer Systems Engineering of Monash University, Melbourne, Australia. He has completed his Ph.D. degree at Swinburne University of Technology, and B.Sc., M.Sc. degrees at Middle East Technical University, Ankara, Turkey. He has lectured at Swinburne University of Technology, Melbourne, Australia for 8 years. His recent research interests are distributed algorithms for self-organization in wireless networks, application of intelligent techniques for multiservice networks as complex, distributed systems.

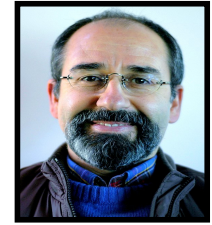

**Milosh V. Ivanovich** fills the role of Senior Emerging Technology Specialist within the Chief Technology Office of Telstra, and is an Honorary Research Fellow at Melbourne and Monash Universities in Australia. A Senior Member of IEEE, Milosh's interests lie in queuing theory, teletraffic modeling, performance analysis of wireless networks, and the study and enhancement of TCP/IP in hybrid fixed/wireless environments. Milosh obtained a B.E. (1st class Hons.) in Electrical and Computer Systems Engineering (1995), a Master of Computing (1996) and a Ph.D. in Information Technology (1998), all at Monash University.

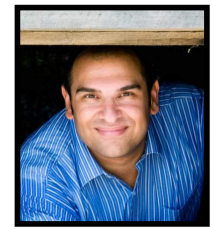

**Paul G. Fitzpatrick** completed his Bachelor Degree in Electrical Engineering at Caulfield Institute of Technology, Melbourne in 1979 and his PhD in Electrical Engineering at Swinburne University, Melbourne in 1997 in the teletraffic performance of hierarchical wireless networks. Paul has over 30 years of experience working in the telecommunications industry and academia, including 15 years at Telstra Research Laboratories working on 2G, 3G and 4G wireless networks. His research interests focus on teletraffic modeling, quality of service, TCP performance modeling and analysis of telecommunication networks.

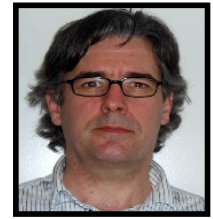

**Jonathan C. Li** received the B.E.in electrical and Electronic Engineering in 2001, B.Sc. degree in Computer Science and Information Technology Systems in 1999 from the University of Western Australia, and Ph.D. in Telecommunication from the University of Melbourne in 2010. He is currently a member of the academic staff at the Department of Electrical and Computer Systems Engineering of Monash University, Melbourne, Australia. His research interests are optical performance monitoring, routing in all-optical networks, network simulation and modeling, and Wireless TCP/IP optimization.

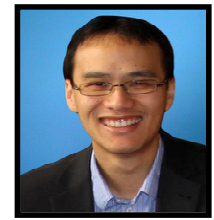